\documentclass[aps,twocolumn,amsmath,amssymb,showpacs,floatfix,prb]{revtex4}
\usepackage{graphicx}

\newcommand{\etal}{{\it et al.}}

\begin{document}

\title{High Frequency Behavior of the Infrared Conductivity of Cuprates}
\author{M. R. Norman$^1$ and A. V. Chubukov$^2$}
\affiliation{
$^1$Materials Science Division, Argonne National Laboratory, Argonne,
Illinois 60439}
\affiliation{$^2$Department of Physics, University of Wisconsin, Madison, WI
53706}
\date{\today}
\begin{abstract}
We analyze recent infrared conductivity data in the 
normal state of the cuprates.
 We find that the high frequency behavior, 
which has been
suggested as evidence for quantum critical scaling, is generally
 characteristic of electrons interacting with a broad spectrum of bosons. 
 From explicit calculations, we
find a frequency exponent for the modulus of the conductivity, and a phase angle, in good
agreement with experiment. 
The data indicate an upper cut-off of the boson spectrum of 
order 300 meV.  This implies that the 
 bosons  are of
 electronic origin
 rather than phonons.
\end{abstract}

\maketitle

Infrared conductivity has proven to be a powerful probe of the electronic 
 degrees of
freedom of the cuprates \cite{BASOV}.  It has the advantage of being 
 bulk sensitive,
and yields useful information over a wide range of energies.  Of particular
interest is a generalized Drude analysis of the data, which provides information on
the optical analogue of the fermion self-energy \cite{PUCH}.  Most data indicate a
linear frequency dependence of the imaginary part of the optical self-energy (i.e., $1/\tau$)
up to very high energies.  Such behavior is characteristic of a marginal Fermi liquid \cite{MFL}.
In some data \cite{NICOLE},
 this trend persists up to the plasma frequency (1 eV).

Recently, van der Marel and collaborators \cite{VDM} have obtained somewhat different
behavior at high frequencies for $1/\tau$, showing a tendency to saturate \cite{BORIS}
above 0.5 eV.
 They have also found that 
in a wide frequency range (125 meV to 
900 meV), both real ($\sigma_1$)  
 and imaginary ($\sigma_2$) parts of the optical 
conductivity are described by the same power law ($\omega^{\alpha}$) with 
an exponent -0.65 (and an 
 associated phase angle $\phi = \tan^{-1}(\sigma_2/\sigma_1)$ of 60$^\circ$).
The same behavior was  observed earlier by El Azrak \etal \cite{NICOLE}.
  The exponent and phase angle in this frequency range are roughly temperature independent, 
  and have been suggested to be indicative of
quantum critical scaling \cite{VDM}.

In this paper, we analyze  the frequency dependent optical data using a model
based on electrons interacting with a broad spectrum of bosons.  We find that the essential results
mentioned above are captured by this analysis, 
indicating that the observed behavior is generic for  interacting electrons. 
We show that the power-law behavior of the conductivity $\sigma$ is not 
indicative of quantum-critical scaling, but rather a consequence of the 
flattening of the fermionic self-energy at high frequencies.
Based on our analysis, we find evidence for an upper cut-off scale of 
the boson spectrum of about 300 meV in the cuprates. This is
 consistent with the assumed value in the
marginal Fermi liquid phenomenology \cite{MFL}, and also 
with the measured width of the spin-fluctuation spectrum \cite{HAYDEN}.

The Kubo expression for the optical conductivity can be written as \cite{ALLEN}

\begin{equation}
\sigma(\omega) = \frac{\omega_{pl}^2}{4\pi} \int d\epsilon \frac{f(\epsilon)-f(\omega+\epsilon)}
{i\omega}\frac{1}{\omega -\Sigma^*(\epsilon)+\Sigma(\omega+\epsilon)}
\label{1}
\end{equation}
where $\omega_{pl}$ is the bare plasma frequency and
$\Sigma$ is the 
retarded
fermion self-energy 
(in this paper we ignore any momentum dependent effects).
Within the same approximation, 
the fermion self-energy can be expressed as
\begin{equation}
\Sigma(\omega) = \int \frac{d\Omega}{\pi} \int d\epsilon \alpha^2F(\Omega) \frac{n_B(\Omega)
+f(\epsilon)}{\omega-\epsilon+\Omega + i \delta}
\end{equation}
with $n_B$ the Bose function and $f$ the Fermi function.
$\alpha^2F$ is the boson spectral function multiplied by the square of the coupling
strength to the fermions and the fermion density of states (and thus is a dimensionless quantity).
For T=0, the imaginary part of the self-energy becomes
\begin{equation}
Im\Sigma(\omega) = \int^{\omega}_0 d\Omega \alpha^2F(\Omega)
\end{equation}
The real part can be obtained by Kramers-Kronig transformation.
From these expressions, one can easily calculate the real 
 and imaginary 
 parts of the conductivity and the phase angle.

Alternately, one can examine the generalized Drude expression for the conductivity \cite{PUCH}
\begin{equation}
\sigma(\omega) = \frac{\omega_{pl}^2}{4\pi}\frac{1}{1/\tau(\omega)-i\omega m^*(\omega)}
\end{equation}
An approximation for $1/\tau$ (i.e., $Im\Sigma_{opt}$) can be obtained
\cite{ALLEN} by expanding the denominator 
of Eq.~1 to lowest order, integrating over frequency, and then inverting the result, again using the 
lowest order expansion, to get $\sigma^{-1} (\omega)$.
We call this the Allen approximation. Within this approximation,
\begin{equation}
\frac{1}{\tau(\omega)} = \frac{2}{\omega} \int_0^\omega {\textrm Im} 
\Sigma (\Omega) d\Omega = 
  \frac{2}{\omega} \int^{\omega}_0 d\Omega (\omega-\Omega)\alpha^2F(\Omega)
\end{equation}
The optical mass, $m^*(\omega) = 1 + Re\Sigma_{opt}/\omega$
 can then be obtained by Kramers-Kronig.

Surprisingly, we have found that for a broad spectrum of bosons, $1/\tau$ determined exactly from
Eq.~1 matches the Allen approximation to a high precision over the entire
frequency range~\cite{comm1}, and therefore for the
purposes of this paper,  either expression can be used interchangeably.
Because of this, we can easily perform
a finite T calculation by using the finite T version of the Allen approximation derived 
in Ref.~\onlinecite{SHULGA}
\begin{eqnarray}
1/\tau(\omega) = \frac{1}{\omega} \int^{\infty}_0 d\Omega \alpha^2F(\Omega)
[2\omega \coth(\frac{\Omega}{2T}) \nonumber \\
- (\omega +\Omega) \coth(\frac{\omega+\Omega}{2T})
+ (\omega -\Omega) \coth(\frac{\omega-\Omega}{2T})]
\end{eqnarray}

In addition, there is the impurity contribution.  We assume an energy independent fermion density
of states,  so that the impurity contribution to $Im\Sigma$ is a constant
which we denote as $\Gamma_i$, thus the contribution to $1/\tau$ is $2\Gamma_i$
(with no change to the optical mass).

Since we are addressing data in the normal state, we consider electrons interacting with a broad
spectrum of bosons.  We have considered two models.
  First a Lorentzian spectrum 
\begin{equation}
\alpha^2 F (\Omega) = {\textrm Im} \frac{\Gamma}{\gamma - i\Omega},
\label{lor}
\end{equation}
 which has been introduced in the context of spin-fluctuation exchange\cite{MMP,HWANG}, 
 and used for the charge propagator as well \cite{GRILLI}.
In our case we will also incorporate a high frequency cut-off into Eq.~\ref{lor}.
Second,  a gapped marginal Fermi 
liquid (MFL) \cite{LITTLE} with
\begin{equation}
\alpha^2 F (\Omega) = {\textrm Im} \frac{1}{\pi}\frac{\Gamma}{\omega_2 - \omega_1}
~\ln{\frac{\omega^2_2 - (\Omega + i \delta)^2}{\omega^2_1 - (\Omega + i \delta)^2}}
\label{mar}
\end{equation}
This model yields a flat $\alpha^2 F (\Omega)$ between  
lower ($\omega_1$) and upper ($\omega_2$) cut-offs.
Both spectra give similar results.

We start with the MFL model.  At T=0, the imaginary part of the fermion self-energy is
\begin{eqnarray}
Im\Sigma = & \Gamma_i & \omega < \omega_1 \nonumber \\
 & \Gamma_i + \Gamma \frac{\omega - \omega_1}{\omega_2-\omega_1} & 
 \omega_1 < \omega < \omega_2 \nonumber \\
 & \Gamma_i + \Gamma & \omega_2 < \omega
\end{eqnarray}
where $\Gamma$ is the frequency integrated spectral weight for $\alpha^2F$.
The real part of the self-energy is easily determined by Kramers-Kronig
\begin{eqnarray}
Re\Sigma = \frac{-\Gamma}{\pi}(
\frac{\omega-\omega_1}{\omega_2-\omega_1}\ln\frac{|\omega-\omega_2|}
{|\omega-\omega_1|} \nonumber \\
+\frac{\omega+\omega_1}{\omega_2-\omega_1}\ln\frac{|\omega+\omega_2|}
{|\omega+\omega_1|}
+\ln\frac{|\omega+\omega_2|}{|\omega-\omega_2|})
\label{si_step}
\end{eqnarray}
The 
 expressions for the real and imaginary parts of the optical self-energy in
the Allen approximation are
\begin{eqnarray}
1/\tau = & 2\Gamma_i & \omega < \omega_1 \nonumber \\
 & 2\Gamma_i + \frac{\Gamma}{\omega} \frac{(\omega - \omega_1)^2}{\omega_2-\omega_1}
 & \omega_1 < \omega < \omega_2 \nonumber \\
 & 2\Gamma_i + 2\Gamma - \frac{\Gamma}{\omega}(\omega_2+\omega_1) & \omega_2 < \omega
\label{tau_step}
\end{eqnarray}
\begin{eqnarray}
Re\Sigma_{opt} = \frac{\Gamma}{\pi \omega (\omega_2-\omega_1)}
[(\omega-\omega_2)^2\ln|\omega-\omega_2| \nonumber \\
 +(\omega+\omega_2)^2\ln|\omega+\omega_2|
- (\omega-\omega_1)^2\ln|\omega-\omega_1|  \nonumber \\
- (\omega+\omega_1)^2\ln|\omega+\omega_1| 
-2\omega_2^2\ln\omega_2 +2\omega_1^2\ln\omega_1]
\label{m_step}
\end{eqnarray}

For finite T, 
 a simple analytic expression for $1/\tau (\omega)$ can be obtained
 only at frequencies $\omega> \omega_2$:
%
\begin{equation}
1/\tau_{high} = 2\Gamma_i + \frac{\Gamma}{\omega_2-\omega_1}
(4T\ln\frac{\sinh\frac{\omega_2}{2T}}{\sinh\frac{\omega_1}{2T}} - \frac{\omega_2^2-\omega_1^2}{\omega})
\end{equation}
We determine $1/\tau$ by numerical integration of Eq.~6  at frequencies below 0.5 eV, and use
Eq.~13 above this frequency.  The optical mass is then determined by numerical Kramers-Kronig.

We have also examined Lorentzian models with either a hard or soft cut-off.
To impose a hard cut-off, we cut $\alpha^2 F (\Omega)$ in Eq.~\ref{lor} at some $\omega_c >> \gamma$; to impose a soft cut-off, we add 
a quadratic frequency term to the denominator of Eq.~\ref{lor}. 
We obtained similar results in both cases. For brevity, 
we present only the results for the hard cut-off case.
The self-energy at $T=0$ is given by
\begin{eqnarray*}
Im\Sigma = & \Gamma_i  + \frac{\Gamma}{2} \ln{\frac{\omega^2 + \gamma^2}{\gamma^2}} & \omega < \omega_c \nonumber \\
 & \Gamma_i + \frac{\Gamma}{2} \ln{\frac{\omega_c^2 + \gamma^2}{\gamma^2}} & \omega > \omega_c 
\end{eqnarray*}
 The expressions for the imaginary part of the optical conductivity  
 in the Allen approximation is 
\begin{eqnarray*}
1/\tau = & 2\Gamma_i + \Gamma(\ln\frac{\omega^2+\gamma^2}{\gamma^2}
-2+\frac{2\gamma}{\omega}
\tan^{-1}\frac{\omega}{\gamma}) & \omega < \omega_c \nonumber \\
 & 2\Gamma_i + \Gamma(\ln\frac{\omega_c^2+\gamma^2}{\gamma^2}
-2\frac{\omega_c}{\omega}+\frac{2\gamma}{\omega}
\tan^{-1}\frac{\omega_c}{\gamma}) & \omega > \omega_c 
\end{eqnarray*}


We  start with some general observations.
The behavior of the optical self-energy is similar in the two models.
 $Re \Sigma_{opt}$ is initially linear for small frequencies, then bends over, passing
 through a maximum near the cut-off, with a decay at higher frequencies.
 $1/\tau$ is linear  (except at the lowest frequencies) up to the cut-off.  Beyond this, it continues
 to rise, but much more slowly.  It should be noted that in the linear regime, the slopes of  
 $Im\Sigma$ and $1/\tau$ are almost identical, unlike the impurity contribution which differs by a factor of two.   This can be understood quite simply from Eq.~3.
Therefore, we expect that the energy derivative of the scattering rates from 
photoemission and optics should coincide, which is indeed the case \cite{ADAM}. 

ARPES has yet to address the question of saturation at high frequencies, but optics has.
Earlier studies indicated that the linear frequency dependence of $1/\tau$ persists to energies
of order 1 eV, but recently van der Marel \etal \cite{VDM} have seen evidence for saturation.
The difference from earlier work has to do with the choice of $\epsilon_{\infty}$ and $\omega_{pl}$ 
($1/\tau$ and $m^*$ are related to the 
dielectric function $\epsilon$ as  $\omega^2m^* + i\omega/\tau = \omega_{pl}^2/(
\epsilon_{\infty}-\epsilon)$)~\cite{comm3}.  
From an analysis of their data, these authors 
 gave
 evidence for quantum critical scaling \cite{VDM}.
In fact, the tendency for saturation in $1/\tau$, as noted above, is actually indicative of being 
above the quantum-critical regime.

\begin{figure}
\centerline{\includegraphics[width=3.4in]{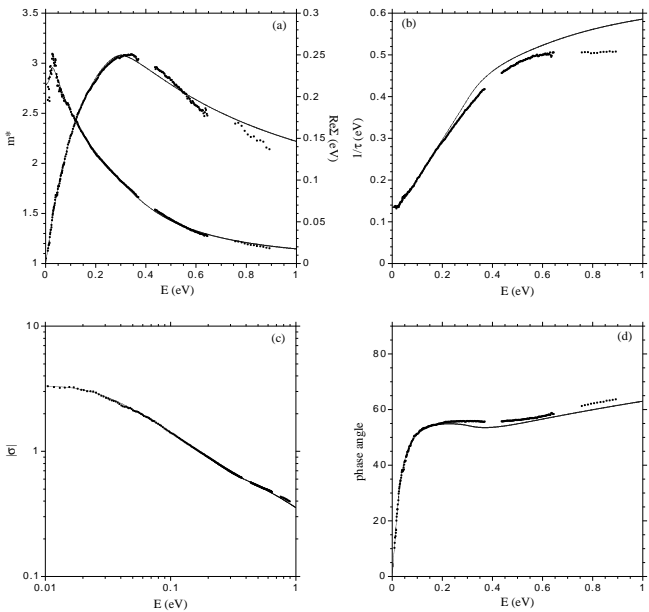}}
\caption{Fit of the MFL model to T=260K data of van der Marel \etal on optimal doped 
Bi$_2$Sr$_2$Ca$_{0.92}$Y$_{0.08}$Cu$_2$O$_{8+\delta}$ \cite{VDM}.  Parameters are (meV)
$\Gamma_i$=67.5, $\Gamma$=270.5, $\omega_1$=15.5, and $\omega_2$=301.  Plotted are
(a) the optical mass and $Re\Sigma_{opt}$, (b) 1/$\tau$, (c) the modulus of $\sigma$, and (d) the
phase angle, $\tan^{-1}(\sigma_2/\sigma_1)$. }
\label{fig1}
\end{figure}

\begin{figure}
\centerline{\includegraphics[width=3.4in]{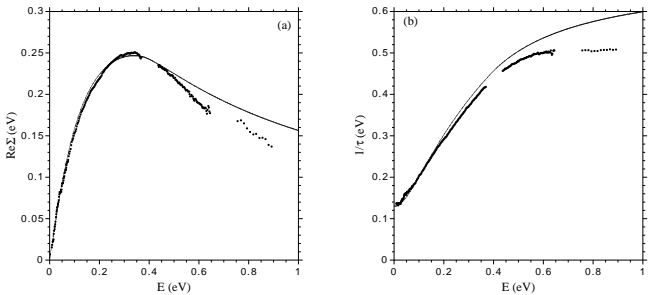}}
\caption{Fit of the Lorentzian model to T=260K data of van der Marel \etal \cite{VDM}.  Parameters are (meV)
$\Gamma_i$=65, $\Gamma$=172, $\gamma$=75, and $\omega_c$=380. }
\label{fig2}
\end{figure}

\begin{figure}
\centerline{\includegraphics[width=3.4in]{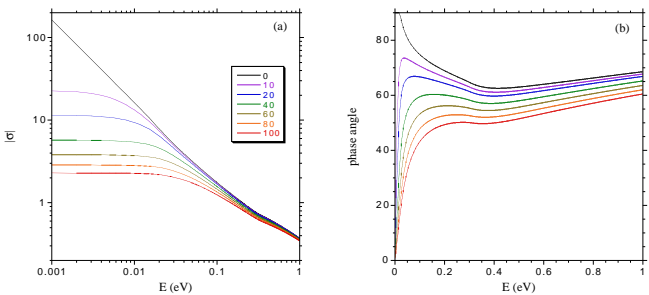}}
\caption{Variation of the MFL model results with impurity scattering strength $\Gamma_i$ (meV).  
Same parameters as Fig.~1.}
\label{fig3}
\end{figure}

\begin{figure}
\centerline{\includegraphics[width=3.4in]{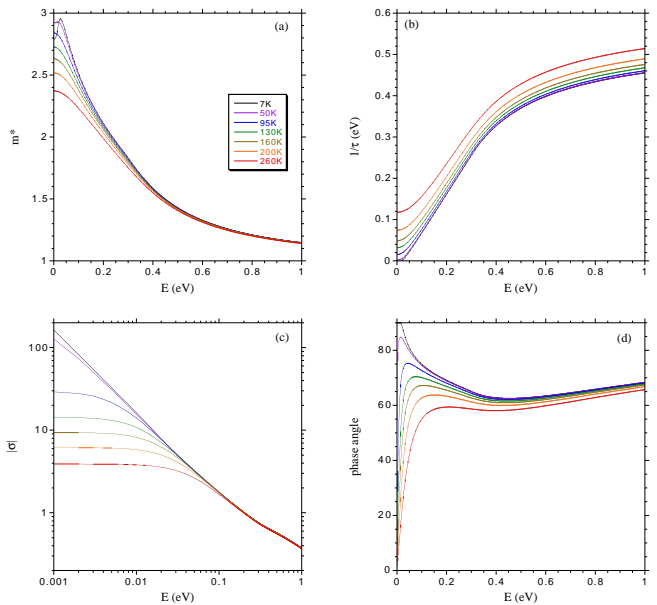}}
\caption{Variation of the MFL model results with temperature.  Same parameters as Fig.~1, except $\Gamma_i$=0.}
\label{fig4}
\end{figure}

To see this point more clearly, we show in Fig.~1 a fit to their data using the MFL model.
We chose to fit the highest temperature data (260K) as the data at lower temperatures give evidence
for a pseudogap effect, which is known from ARPES to be present up to around 200K for optimal
doped samples \cite{JC99}.  The fit was performed to the optical mass using Eq.~\ref{m_step}.
The T=0 expression was used as it is analytic and thus can be employed in any non-linear fitting routine.

Some remarks are in order.  First, the low frequency cut-off is evident as a peak in the optical mass
at  about 15 meV.  The high frequency cut-off is evident as a peak in $Re\Sigma_{opt}$ at about 
300 meV.
Therefore, even without the fit, these values can be read off directly from the data.
We should remark that the low frequency cut-off is not that important (it simply assures that
the optical mass
does not diverge at low frequencies), and thus the fit to $Re\Sigma_{opt}$ is essentially a two
parameter one ($\Gamma$ and $\omega_2$).
Second, the upper cut-off is also visible where $1/\tau$ deviates from linear behavior.  We note
that the mismatch between the fit and data for $1/\tau$ at high frequencies can be compensated 
by a small shift 
in the assumed value of $\epsilon_{\infty}$ and so is not a serious issue.  Third, the fit gives an 
excellent reproduction of the modulus of $\sigma$, and in particular the 
exponent value of -0.65.  Therefore, the fact that this value is fractional does not necessarily imply quantum
critical physics with a sub-linear exponent.  Moreover, we note that the phase angle is well reproduced
by the fit.

We have obtained similar results by fitting to a Lorentzian with a high frequency cut-off (Fig.~2).  
Therefore,
the high frequency data should not be taken as being dependent on having a marginal Fermi liquid
bosonic spectrum, but rather is a generic feature of electrons interacting with a broad spectrum
of bosons.  This is evident as well from the work of Hwang \etal \cite{HWANG2}.

In Figs.~1 and 2, a rather large value is needed for $\Gamma_i$ to fit the zero frequency limit of $1/\tau$.
As is obvious from the linear T dependence of the resistivity, most of this term is actually inelastic.
To examine this in more detail, we show the variation of the modulus of
the conductivity and the phase angle as a function of $\Gamma_i$ (Fig.~3) and T (Fig.~4).
Both variations are similar, and reproduce well the experimental variation with temperature \cite{VDM}.  
Note that the phase angle is always zero at zero energy unless
T=0, $\Gamma_i$=0,  where it becomes 90 degrees.

What are the implications of this work?  First, 
 we see that the apparent scaling behavior over a wide frequency range is
 actually unrelated to quantum criticality and is just the consequence
 of the flattening of $1/\tau$, 
 accompanied by a decrease in $Re \Sigma_{opt}$. Second, we see that the
 behavior of the single particle and optical self-energies is very similar 
 for the  marginal Fermi liquid phenomenology, and the Lorentzian model used in microscopic fermion-boson theories. Third,   
the data give strong evidence for an upper cut-off of
the boson spectrum of around 300 meV.
A cut-off of this scale was suggested in the original marginal Fermi liquid 
phenomenology \cite{MFL}.  Such a large energy scale would imply that the source of the boson
spectrum is collective electronic excitations rather than phonons.
We note that inelastic neutron scattering data show magnetic spectral weight up to this energy 
scale \cite{HAYDEN}, and thus spin fluctuations are a natural explanation
for the boson spectrum.  This would be in support of a magnetic origin for cuprate superconductivity.

We would like to thank Tom Timusk, Dmitri Basov, Chris Homes,
Nicole Bontemps, Andres Santander-Syro, and Ricardo Lobo for many discussions
concerning the infrared data.  We especially thank Dirk van der Marel and Hajo
Molegraaf for many exchanges concerning their results.
The work by MRN was supported by the U. S. Dept. of Energy, Office of Science,
under Contract No. W-31-109-ENG-38. A.C. acknowledges support by NSF-DMR 0240238 and would like to thank ANL for hospitality 
 during his visit when this work was initiated.


\begin{thebibliography}{99}
\bibitem{comm1} The validity of the Allen approximation is the result of the interplay between the two approximate procedures involved: the expansion in $\Sigma$ in Eq.~\protect\ref{1}, and the inversion of the result to get $\sigma^{-1} (\omega)$. The extension of the procedure to second order in $\Sigma$ yields, at $T=0$, $\sigma^{-1} (\omega) = -(4i \pi \omega/\omega^2_{pl}) 
(1 + (2/\omega^2) \int_0^\omega \Sigma (\Omega) d\Omega + (4/\omega^4) (\int_0^\omega \Sigma 
(\Omega) d\Omega)^2$

$ -(2/\omega^3) \int_0^\omega \Sigma(\Omega) (\Sigma (\Omega) + \Sigma (\omega-\Omega))d \Omega)$. In principle, the 
 $O(\Sigma^2)$ contribution can be large at strong coupling. In practice, the
 two last terms in $\sigma^{-1} (\omega)$ almost cancel each other. 
For a generic non-Fermi-liquid form of 
$\Sigma (\Omega) = \omega^\alpha (i\omega_0)^{1-\alpha}$ and $\omega \sim \omega_0$, the sum of the last two terms is less than a percent of the $O(\Sigma)$ term for all $\alpha \leq 1$.
\bibitem{comm3}
 From fits to ellipsometry data in the ultraviolet range 
(1 - 8 eV), the authors of Ref.~\protect\onlinecite{VDM} determine a value of 
$\epsilon_{\infty}$ appropriate for the infrared range of interest (0 - 1 eV).  They then choose the
bare plasma frequency $\omega_{pl}$ in such a way that the optical mass approaches one  at
high frequencies.  The resulting value (2.4 eV) is comparable to the band theory one.
\end{thebibliography}
\end{document}